\documentclass[a4paper,11pt,centertags,reqno]{amsart}

\usepackage{letltxmacro}        						
\usepackage{ifthen}													
\usepackage[headings]{fullpage} 
\usepackage[nodisplayskipstretch]{setspace}						
\usepackage[inline]{enumitem}   							
\usepackage{array}														
\usepackage{fancybox}													    
\usepackage{graphicx}           					
\usepackage{amsmath}	        	
\usepackage{amsthm}							
\usepackage{amssymb}	
\usepackage{relsize}		
\usepackage{mathtools}					
\usepackage{mathrsfs}												
\usepackage{yhmath}							
\usepackage{accents}						       
\usepackage{xfrac}             
\usepackage{url}
\usepackage[all]{xy}						
\usepackage[normalem]{ulem}
\usepackage[round,comma,authoryear,sort&compress]{natbib} 
\usepackage{lmodern}
\usepackage{anyfontsize}
\PassOptionsToPackage{hyphens}{url}
\usepackage[colorlinks,citecolor=blue,urlcolor=blue,unicode,verbose]{hyperref}
\usepackage{color}
\usepackage{epigraph}
\usepackage{xr}
\usepackage{tikz}
\usepackage{doi}
\DeclareFontFamily{U}{mathx}{\hyphenchar\font45}
\DeclareFontShape{U}{mathx}{m}{n}{
      <5> <6> <7> <8> <9> <10>
      <10.95> <12> <14.4> <17.28> <20.74> <24.88>
      mathx10
      }{}
\DeclareSymbolFont{mathx}{U}{mathx}{m}{n}
\DeclareFontSubstitution{U}{mathx}{m}{n}
\DeclareMathAccent{\widecheck}{0}{mathx}{"71}
\DeclareMathAccent{\wideparen}{0}{mathx}{"75}

\makeatletter
\newcommand*\rel@kern[1]{\kern#1\dimexpr\macc@kerna}
\newcommand*\widebar[1]{%
  \begingroup
  \def\mathaccent##1##2{%
    \rel@kern{0.8}%
    \overline{\rel@kern{-0.8}\macc@nucleus\rel@kern{0.2}}%
    \rel@kern{-0.2}%
  }%
  \macc@depth\@ne
  \let\math@bgroup\@empty \let\math@egroup\macc@set@skewchar
  \mathsurround\z@ \frozen@everymath{\mathgroup\macc@group\relax}%
  \macc@set@skewchar\relax
  \let\mathaccentV\macc@nested@a
  \macc@nested@a\relax111{#1}%
  \endgroup
}
\makeatother

\newcommand{\N}{\mathbb{N}}

\newcommand{\R}{\mathbb{R}}

\newcommand{\M}{\mathsf{M}}

\newcommand{\SR}{\mathrm{SR}}
\newcommand{\HR}{\mathrm{HR}}

\newcommand{\MSR}{\widebar{\mathrm{SR}}}
\newcommand{\MHR}{\widebar{\mathrm{HR}}}

\newcommand{\E}{\textsf{\upshape E}}

\newcommand{\indicator}[1]{\mathbf{1}_{#1}}

\DeclarePairedDelimiterX\br[1]{[}{]}{#1}

\DeclarePairedDelimiterX\set[2]\{\}{#1\::\:#2}

\makeatletter
\let\oldr@@t\r@@t
\def\r@@t#1#2{%
  \setbox0=\hbox{$\oldr@@t#1{#2\,}$}\dimen0=\ht0
  \advance\dimen0-0.2\ht0
  \setbox2=\hbox{\vrule height\ht0 depth -\dimen0}%
  {\box0\lower0.4pt\box2}}
\LetLtxMacro{\oldsqrt}{\sqrt}
\renewcommand*{\sqrt}[2][\ ]{\oldsqrt[#1]{#2}}
\makeatother

\theoremstyle{plain}
\newtheorem{theorem}{Theorem}

\theoremstyle{definition}

\newtheorem{example}[theorem]{Example}

\theoremstyle{remark}

\numberwithin{theorem}{section}
\numberwithin{equation}{section}
\numberwithin{figure}{section}
\numberwithin{table}{section}

\graphicspath{{Figures/}}
\bibpunct{(}{)}{,}{a}{}{,}

\allowdisplaybreaks[4]
\makeatletter
\@namedef{subjclassname@2010}{MSC {[2020]}}
\makeatother

\begin{document}
\title[The Hansen Ratio]
{The Hansen Ratio in Mean--Variance Portfolio Theory}

\author{Ale\v{s} \v{C}ern\'{y}}

\address{Ale\v{s} \v{C}ern\'{y}\\
  Business School\\
	City, University of London}

\email{ales.cerny.1@city.ac.uk}

\thanks{I would like to thank Christoph Czichowsky for helpful discussions.}   


\renewcommand{\keywordsname}{Keywords}


\date{\today}

\begin{abstract} 
It is shown that the ratio between the mean and the $L^2$--norm leads to a particularly parsimonious description of the mean--variance efficient frontier and the dual pricing kernel restrictions known as the Hansen--Jagannathan (HJ) bounds. Because this ratio has not appeared in economic theory previously, it seems appropriate to name it the Hansen ratio. The initial treatment of the mean--variance theory via the Hansen ratio is extended in two directions, to monotone mean--variance preferences and to arbitrary Hilbert space setting. A multiperiod example with IID returns is also discussed. 
\end{abstract}

\maketitle

\section{Introduction}\label{sect: intro}

\citet{roy.52} gave the first formula for the efficient mean-variance frontier in a one-period market spanned by a finite number of assets. Since then a concerted effort has been made to compute the mean-variance frontier in a dynamic setting, typically in the context of quadratic hedging; see \citet{li.ng.00}, \citet{bertsimas.al.01}, and \citet{lim.04,lim.05}, for example. The results in these studies are explicit; yet despite or perhaps because of this, one gets no closer to a good economic understanding of the underlying principles that drive them.

The literature also features a parallel stream in an abstract market setting with possibly infinitely many assets where geometry plays an important role. The geometric approach starts with \citet{chamberlain.rothschild.83}. Its objects take a more explicit form in \citet{hansen.richard.87} who identify two important portfolios that fully describe the efficient frontier (the portfolios $Y$ and $X$ below).%
\footnote{An accessible account of this literature appears in \citet[Chapter~5]{cochrane.01}.}
 
In this paper, we rediscover and extend the seminal contributions of \citet{hansen.richard.87} and \citet{hansen.jagannathan.91} in the context of utility maximization. This naturally leads to the ratio of mean to $L^2$--norm: the Hansen ratio from the title. The ratio plays an important role in the description of the efficient frontier (Subsection~\ref{SS:3}); it features prominently in the Hansen--Jagannathan bound (Subsection~\ref{SS:HJbound}); and last but not least, it is instrumental in converting the one-period efficient frontier into the dynamically efficient frontier (Section~\ref{S:5} and Appendix~\ref{S:A}). The expected utility approach also yields a clean economic intepretation of the Hansen--Jagannathan bound under positivity constraints in terms of the monotone Hansen ratio/monotone Sharpe ratio (Section~\ref{S:MHR}). In Section~\ref{S:4}, the Hilbert space generalization of the $L^2$ theory is illustrated on an example from \citet{cochrane.14}. Section~\ref{S:6} concludes.

\section{\texorpdfstring{The $L^2$}{L2} theory}
Denote by $L^2$ the collection of all random variables with finite second moment on some fixed probability space. Let $\M$ be a closed linear subspace of $L^2$ and $\pi$ a continuous linear functional on $\M$. We think of elements of $\M$ as terminal wealths $W$ of traded positions whose initial price is given by $\pi(W)$. Denote by $\M(c)$ all traded wealth distributions available at cost $c\in\R$,
$$ \M(c) =\left\{W\in\M:\pi(W)=c\right\},$$
and assume $\M(1)$ is not empty. The elements of $\M(1)$ are commonly known as the fully invested portfolios. With \citet{cochrane.01}, we will refer to the elements of $\M(0)$ as zero-cost portfolios.%
\footnote{The role of $\pi$ is significant only to the extent that $\M(1)$ is a closed subset of $L^2$, that zero position is not an element of $\M(1)$ (as linearity yields $\pi(0)=0$), and that we have $\M(0)=\M(1)-\M(1)$. One could therefore use a closed affine subspace $\M(1)$ not containing zero as the primitive input.}

For each $W$ with finite second moment we write $\mu_W=E[W]$ for the mean, $\omega^2_W=E[W^2]$ for the second non-central moment, and $\sigma^2_W=E[(W-\mu_W)^2]$ for the variance. Observe that $\omega_W = \|W\|$ is the $L^2$--norm of $W$. Observe also that $\mu_W$, $\sigma_W$, and $\omega_W$ are tied together through the relationship $\sigma_W^2=\omega_W^2-\mu_W^2$. The Hansen ratio, 
$$\HR_W = \frac{\mu_W}{\omega_W},$$ 
therefore satisfies the inequality
\begin{equation}\label{eq:risk-free bound} 
\HR_W^2\leq 1.
\end{equation}
Furthermore, $\HR^2_W=1$ if and only if $W$ is risk-free.  

The portfolio theory is concerned with two questions: how to describe all pricing rules on $L^2$ that are consistent with $\pi$ and how to identify efficient portfolios in $\M(1)$. It turns out there is a special portfolio in $\M(1)$ that forms an important part of the answer in both directions.

\subsection{Special fully invested portfolio \texorpdfstring{$Y$}{Y}}\label{SS:Y}
Under our assumptions, there is a unique portfolio in $\M(1)$ orthogonal to $\M(0)$. Let us denote this portfolio by $Y$. As 
\begin{equation}\label{eq:M1 decomp}
\M(1) = Y \oplus^{\mathsmaller{\bot}} \M(0),
\end{equation}
we observe that $Y$ is also the unique element of $\M(1)$ with the smallest $L^2$ norm. This means $Y$ is efficient in the sense of having the smallest variance at the fixed mean $\mu_Y$, in view of $\sigma^2_Y=\omega^2_Y -\mu^2_Y$ and the minimality of $\omega^2_Y$ among all fully invested portfolios. The orthogonality \eqref{eq:M1 decomp} yields that the linear functional
\begin{equation}\label{eq:piY}
\pi^Y(W)=\frac{\E[YW]}{\omega^2_Y},\qquad W\in L^2,
\end{equation}
correctly prices all positions in the marketed subspace $\M$, inasmuch $\pi^Y$ correctly prices $Y$ at 1 and all elements of $\M(0)$ at zero.
The special pricing kernel $\pi^Y$ goes back to at least \citet{chamberlain.rothschild.83}. The portfolio $Y$ appears explicitly in \citet[Eq.~3.6]{hansen.richard.87}.

\subsection{Special zero-cost portfolio \texorpdfstring{$X$}{X}}\label{SS:X}
Let us now consider the quadratic utility function 
\begin{equation}\label{eq:U}
U(x) = x-x^2/2.
\end{equation}
The expected quadratic utility preference reads
$$ u(W)=\E[U(W)]=\mu_W - \frac{1}{2}\omega_W^2,\qquad W\in L^2.$$
One easily verifies that the maximal utility from an optimally scaled investment $W$ equals
\begin{equation}\label{eq:u and HR}
 \max_{\alpha\in\R}u(\alpha W) = \frac{1}{2}\HR^2_W.
\end{equation}

Denote by $X\in\M(0)$ the zero-cost portfolio with the highest expected utility. After completing $U$ to a square,
$$U(x) = \frac{1}{2} - \frac{1}{2}(1-x)^2,$$
we observe that $X$ is the orthogonal projection of the constant payoff $1$ (which here signifies the bliss point of the utility $U$) onto the zero-cost subspace $\M(0)$.%
\footnote{Because $\M(0)$ is a closed subspace of $L^2$, this establishes the existence of $X$.}
 This implies $1-X$ is orthogonal to all elements of $\M(0)$, in particular to $X$ itself, which yields%
\footnote{By the same token, the Hansen ratio of the residual $1-X$ is complementary to the Hansen ratio of $X$, i.e.,
$$ \mu_{1-X} = \omega^2_{1-X} = \HR^2_{1-X}=1-\HR^2_X.$$}
$$ \mu_X = \omega^2_X = \HR^2_X.$$

As $X$ has the smallest value of $\mu - \omega^2/2$ among all zero-cost portfolios, it must be efficient in $\M(0)$. In particular, for the given mean $\mu_X$, there can be no zero-cost portfolio with the second moment smaller than $\omega^2_X$. It immediately follows that 
\begin{itemize}
\item \emph{all} efficient zero-cost portfolios are a constant multiple of $X$; 
\item $\HR_X$ is the highest Hansen ratio available in $\M(0)$.
\end{itemize}

The performance of zero-cost portfolios is often quoted in terms of their Sharpe ratio. By straightforward calculations, one obtains the following identities/conversions,
\begin{equation}\label{eq:HRSR}
1+\SR_{W}^{2} =\frac{1}{1-\HR_{W}^{2}}; \quad \SR_{W} =\frac{\HR_{W}}{\sqrt{1-\HR_{W}^{2}}}; \quad
\HR_{W} =\frac{\SR_{W}}{\sqrt{1+\SR_{W}^{2}}},\qquad W\in L^2.
\end{equation}%
Assuming one cannot obtain risk-free profit with zero initial outlay, $1\notin \M(0)$, one obtains the following range restrictions for the efficient zero-cost portfolio $X$,
\begin{align*}
0 &\leq \HR_{X}<1, \\
0 &\leq \HR_{X}\leq\SR_{X}<\infty.
\end{align*}

\subsection{The Hansen ratio on the efficient frontier}\label{SS:3}
When two random variables are orthogonal, $E[VW]=0$, their second moments are additive,
$$ \omega^2_{V+W}=\omega^2_V+\omega^2_W.$$
Because the means are additive in any case, we find that the expected quadratic utility of a portfolio of orthogonal investments is additive and hence each component can be optimized separately. From here and from \eqref{eq:u and HR} one immediately draws two conclusions.
\begin{itemize}
\item The squared Hansen ratio of a portfolio of orthogonal investments is subadditive
$$u(\alpha V + \beta W)\le \frac{1}{2}\HR^2_{\alpha V +\beta W}\le \max_{\alpha,\beta\in\R}u(\alpha V + \beta W)
=\frac{1}{2}\left(\HR^2_{V}+\HR^2_{W}\right);$$
\item By changing the risk aversion in the utility function \eqref{eq:U}, one observes that the maximal squared Hansen ratio of a sum of two orthogonal investments is attained even if we fix $\alpha =1$ and vary only $\beta$, provided that $\mu_V\neq 0$. More generally, 
\begin{equation}\label{eq: alpha fixed}
\sup_{\beta\in\R}\HR^2_{V+\beta W}=\HR^2_{V}+\HR^2_{W}.
\end{equation} 
The supremum is attained, except for the case $\mu_V=\HR_V=0$ where it is attained asymptotically with $|\beta|$ going to infinity.
\end{itemize}

Now let us apply these observations to the assets on the efficient frontier. Thanks to \eqref{eq:M1 decomp}, all fully invested portfolios are of the form $Y+W$, where $W$ costs zero. As the means and second moments are additive (the latter due to the orthogonality between $Y$ and $\M(0)$), the portfolio $Y+W$ is efficient if and only if $W$ is efficient in $\M(0)$. 
This shows the efficient frontier is of the form 
$$Y+\lambda X,\qquad\lambda\in\R;$$
a result that goes back to at least \citet[Lemma 3.3]{hansen.richard.87}.

Formula \eqref{eq: alpha fixed} shows that the squared Hansen ratio on the efficient frontier is dominated by $\HR^2_X+\HR^2_Y$; this upper bound is attained for some $\lambda\in\R$ if $\mu_Y\neq 0$ and otherwise it is attained asymptotically for $|\lambda|$ going to infinity.
All this yields a somewhat counter-intuitive result. Because $\HR^2_X+\HR^2_Y$ is the least upper bound of the squared Hansen ratio on the efficient frontier, and because the squared Hansen ratio always satisfies the risk-free bound~\eqref{eq:risk-free bound}, we conclude that 
\begin{equation}\label{eq: BR XY}
\HR^2_X+\HR^2_Y\leq 1.
\end{equation}

The restriction on $Y$ is remarkable; it states the square of the Hansen ratio of $Y$ cannot be an arbitrary number below 1, it must be a number below  $1-\HR_X^2$. Observe that the investment performance of $X$ restricts the possible location of $Y$ in the mean--standard deviation diagram. No such restriction flowing from $X$ applies to the minimum variance portfolio, whose properties we examine next. 

\subsection{The minimum variance fully invested portfolio \texorpdfstring{$Z$}{Z}}\label{SS:Z}
Assume for now that the minimum variance fully invested portfolio exists and denote this portfolio by $Z$. The minimum variance portfolio is evidently efficient because for the given mean $\mu_Z$ there can be no portfolio with standard deviation smaller than $\sigma_Z$. Consequently, there exists $\hat\lambda \in \mathbb{R}$ such that 
\begin{equation}\label{Z by X,Y}
Z=Y+\hat\lambda X.
\end{equation}%

Conversely, let us now identify the minimum variance portfolio on the efficient frontier (thus also establishing existence).
As is well known, variance is the residual sum of squares from an orthogonal projection of the random variable in question onto constant $1$.  As we seek $Z$ of the form \eqref{Z by X,Y}, the minimal variance $\sigma^2_Z$ is therefore the residual sum of squares from an orthogonal projection of $Y$ onto the span of $X$ and $1$,
\begin{equation}\label{eq: Z OLS}
\sigma _{Z}^{2}=\min_{\alpha ,\lambda \in \mathbb{R}}\omega _{Y+\lambda X-\alpha}^{2}
=\min_{\alpha ,\lambda \in \mathbb{R}}\|Y+\lambda X-\alpha \|^{2}
=\min_{\alpha ,\lambda \in \mathbb{R}}\|Y+(\lambda -\alpha)X-\alpha (1-X)\|^{2}.  
\end{equation}%
Because $X$ itself is the projection of $1$ onto $\M(0)$, we have that $1-X$ is orthogonal to $X$. Furthermore, $Y$ too is orthogonal to $X$ so the least squares optimality in \eqref{eq: Z OLS} yields%
\footnote{Equation \eqref{mu_Z} in combination with \eqref{Z by X,Y} gives the remarkable identity
$$Z = Y +\mu_Z X.$$}
\begin{equation}\label{mu_Z}
\mu _{Z}=\hat{\alpha}=\hat{\lambda}.
\end{equation}

The coefficient $\hat{\alpha}$ can now be obtained by regressing $Y$ onto $1-X$ which yields
\begin{equation}\label{alpha^}
\mu_Z =\hat{\alpha} =\frac{\mu_Y}{1-\HR_X^2}.
\end{equation}
From \eqref{eq: Z OLS} it is also immediate that
\begin{equation}\label{sig_Z}
 \sigma_Z^2 = \|Y\|^2-\hat{\alpha}^2\|1-X\|^2=\omega_Y^2 - \frac{\mu_Y^2}{1-\HR_X^2}=\omega_Y^2\left(1-\frac{\HR_Y^2}{1-\HR_X^2}\right).
\end{equation}
Non-negativity of $\sigma_Z^2$ now once again yields the Hansen ratio restriction \eqref{eq: BR XY}.

Observe that the knowledge of any two of the three special portfolios $X$, $Y$, and $Z$ implies the knowledge of the remaining portfolio. It is thus a matter of convenience which two special portfolios one chooses to identify. We will return to this point in the concluding Section~\ref{S:6}. 

\subsection{The Hansen--Jagannathan bound}\label{SS:HJbound}
By following the logic of Subsection~\ref{SS:3}, it is not difficult to see that the `efficient frontier' of pricing kernels is of the form 
\begin{equation}\label{eq:m}
m = \frac{Y}{\omega_Y^2}+\eta V,\qquad\eta\in\R,
\end{equation}
where $V$ is an `efficient' element of $\M^\bot$.  From Subsection~\ref{SS:X} we know $V$ may be taken as the solution of $\min_{W\in\M^\bot}\|1-W\|^2$. This yields $V$ as the residual from the orthogonal projection of $1$ onto the span of $X$ and $Y$ (in fact onto $\M$),
$$ V = 1-X-\frac{\mu_Y}{\omega_Y^2}Y,$$
with  
$$\HR_V^2 = 1 - \HR_X^2 - \HR_Y^2.$$
Variable $V$ is the realized distance from the bliss point of the quadratic utility \eqref{eq:U} for an optimal investment in $\M$ with optimally chosen initial wealth $\mu_Y/\omega_Y^2$. 

We now conclude from \eqref{eq:m} and the arguments in Subsection~\ref{SS:3} that 
\begin{equation}\label{eq:HJbound}
 \HR_m^2 \leq \HR_Y^2 + \HR_V^2 = 1-\HR_X^2.
\end{equation}
This is a previously unpublished version of the Hansen--Jagannathan bound. More commonly, provided $\mu_m\neq 0$, the bound is formulated in terms of variance. By taking reciprocals in \eqref{eq:HJbound} one obtains
$$ \HR_{m}^{-2}-1\geq (1-\HR_X^2)^{-1}-1.$$ 
This then yields, with the help of conversion \eqref{eq:HRSR}, the more commonly encountered formula
\begin{equation} \label{eq:HJbound2}
\frac{\sigma^2_m}{\mu^2_m}\geq \SR_X^2;
\end{equation}
see \citet[Eq. 17]{hansen.jagannathan.91}.

\section{The monotone Hansen ratio and non-negativity constraints}\label{S:MHR}
The next example shows that the Hansen ratio does not preserve the state-wise stochastic dominance ordering. 
\begin{example}\label{E:3.1}
In a three-state model with $W_d=-1\%$, $W_m=1\%$, $W_u=2\%$ and with probabilities $p_d=\frac{1}{6}$, $p_m=\frac{1}{2}$, and $p_d=\frac{1}{3}$ one obtains
$$\HR_W = \frac{1}{\sqrt{2}}, $$
while for $\widetilde W = W + 0.09\indicator{W>0.01}\gneqq W$ one has
\[\pushQED{\qed}
\HR_{\widetilde W} = \frac{4}{\sqrt{41}} < \HR_W.
\qedhere \popQED \]
\end{example}
With \citet{filipovic.kupper.07}, define the monotone Hansen ratio, $\MHR$, as the monotone hull of $\HR$, that is, $\MHR: L^0_+ - L_{+}^2 \to (-\infty,\infty]$ with
\begin{equation}\label{eq:MHR}
\MHR_W = \sup_{\widetilde W\in L_{+}^{0}} \HR_{W-\widetilde W}. 
\end{equation}

Intuitively, the monotonization allows to set aside some non-negative cash amount if this leads to an increase in the Hansen ratio of the remaining wealth. It is shown in \citet{cerny.20} that for $W\in L^0_+ - L_{+}^2$ with positive (possibly infinite) mean and non-zero downside, the supremum in \eqref{eq:MHR} is attained. Furthermore, one has  
\begin{equation}\label{eq:SRm2}
\MHR(W)= \HR((\hat\alpha W)\wedge 1)=\HR(W\wedge \hat{\alpha}^{-1})=\max_{k>0}\HR(\widetilde W\wedge k),
\end{equation}
where $\hat{\alpha}>0$ is the unique solution of 
\begin{equation}\label{eq:FmFOC}
\E[W \indicator{\hat\alpha W\leq 1}] = \hat\alpha\E[W^2 \indicator{\hat\alpha W\leq 1}].
\end{equation}
This shows that one puts away all returns above some fixed threshold $k$ chosen in such a way that the optimal investment with the truncated investment opportunity only just touches but does not reach over the bliss point of the quadratic utility.
\begin{example}\label{E:3.2}
In the setting of Example~\ref{E:3.1}, the optimal value of $k$ in $\max_{k>0}\HR(\widetilde W\wedge k)$ is 2\%, hence
\[\pushQED{\qed}
\MHR^2(\widetilde W) = \HR^2(\widetilde W\wedge 0.02) = \frac{1}{2}.
\qedhere \popQED \]
\end{example}

In effect, the monotone Hansen ratio is obtained from the maximization of monotonized quadratic utility 
$$\widebar{U}(x) = x\wedge 1 - (x\wedge 1)^2/2,$$
so that for $W$ with non-negative mean one obtains, in analogy to \eqref{eq:u and HR}, 
$$\max_{\alpha\in\R} \widebar{u}(\alpha W) = \max_{\alpha\in\R} \E[\widebar{U}(\alpha W)] = \frac{1}{2}\MHR_W^2.$$

By the Fenchel inequality, one obtains for all non-negative pricing kernels $m\in L^2$ and all $W\in\M(0)$
$$\E[\widebar{U}(W)]\leq \min_{\lambda \in\R}\E\left[\frac{1}{2}(1-\lambda m)^2\right].$$
 From here we have the lower bound
$$\sup_{W\in\M(0)}\MHR_W^2\leq 1 - \HR^2_m$$
for all non-negative pricing kernels in $L^2$. Under minimal assumptions this lower bound is tight; see \citet[Theorem~4.3]{biagini.cerny.20}.

Observe that the monotone Sharpe ratio, $\MSR$, is related to $\MHR$ in the same way the standard Sharpe ratio is to $\HR$; see \eqref{eq:HRSR}. The analogon of  the classical Hansen--Jagannathan inequality \eqref{eq:HJbound2} for non-negative pricing kernels thus reads
$$\frac{\sigma_m^2}{\mu_m^2}\geq \sup_{W\in\M(0)}\MSR_W^2.$$

\begin{example}
We remain in the setting of Example~\ref{E:3.1}. A market spanned by the zero-cost investment $\widetilde W$ has the standard squared Sharpe ratio of
$$\SR^2_{\widetilde W} = \frac{1}{1-\HR^2_{\widetilde W}} - 1 = 0.64,$$
 while by Example~\ref{E:3.2}, the square of its monotone Sharpe ratio is 
$$ \MSR^2_{\widetilde W} = \frac{1}{1-\MHR^2_{\widetilde W}} - 1 = 1,$$
which leads to a tightening of the variance bound for non-negative pricing kernels. \qed
\end{example}

\section{The Hilbert space generalization}\label{S:4}
Following \citet{cochrane.14}, let us now consider a Hilbert space with an inner product $\left\langle V,W\right\rangle$.
In the classical mean--variance theory, the Hilbert space is $L^2$ with $\left\langle V,W\right\rangle =E[VW]$. In particular, $\mu_V=\langle V,1\rangle$. 
In the generalized theory the role of $1$ will be played by a special element $I$ that also assumes the role of a `risk-free' payoff. We will set 
\begin{equation*}
\mu_V=\left\langle V,I\right\rangle .
\end{equation*}
To make the analogy complete, we will also require
$$I\notin \M(0);\qquad \left\langle I,I\right\rangle =1.$$

For ease of notation we continue to write 
\begin{align*}
\omega _{V}^{2} &=\left\langle V,V\right\rangle = \| V\|^2;  \\
\sigma _{V}^{2} &=\left\| V-\mu_V I\right\|^2  =\omega _{V}^{2}-\mu _{V}^{2},
\end{align*}%
referring to $\mu _{V}$ as the \textbf{mean}, $\omega _{V}^{2}$ as the \textbf{second moment}, and $\sigma _{V}^{2}$ as the \textbf{variance}, 
even when these objects no longer have such classical interpretation. Observe that $\sigma^2_I=0$ so the payoff $I$ is indeed `risk-free'.

The results in Subsections~\ref{SS:Y}--\ref{SS:HJbound} now translate directly to the Hilbert space setting. One only needs to replace $1-X$ (resp., $1-W$) with $I-X$ (resp., $I-W$) and $\E[YW]$ (resp. $\E[VW]$) with $\left\langle Y,W\right\rangle$ (resp., $\left\langle V,W\right\rangle$).

\begin{example}
\citet{cochrane.14} considers, among others, the Hilbert space of sequences of random variables $V=\{v_n\}_{n\in\N}$ with 
$$ \|V\|^2 = \frac{\beta}{1-\beta}\sum_{n\in\N}\beta^{n}\E[v_n^2]$$ 
for some fixed positive $\beta<1$. The special `risk-free' payoff $I$ is taken to be the constant cash flow $1$ at every date. \qed
\end{example}

\section{The Hansen ratio in dynamic models}\label{S:5}
Consider an $n$--period model with independent, identically distributed (IID) returns. Denote by $\widetilde X$, $\widetilde Y$ the unconditionally optimal portfolios in the dynamic model and by $\{X_t,Y_t\}_{t=1}^n$ their one-period counterparts running from $t-1$ to $t$ for each $t\in\{1,\ldots,n\}$. By the IID assumption the collection $\{X_t,Y_t\}$ is also IID. By dynamic programming one obtains
$$ \widetilde Y = \prod_{t=1}^n Y_t$$
and
$$ 1-\widetilde X - \frac{\mu_{\widetilde Y}}{\omega^2_{\widetilde Y}}\widetilde Y 
= \sum_{j=1}^n \left(\frac{\mu_Y}{\omega_Y^2}\right)^{n-j}\left(1- X_j - \frac{\mu_{Y}}{\omega^2_{Y}}Y_j\right)\prod_{t=j+1}^n Y_t,$$
where an empty product is defined to take the value of 1. 

This yields
\begin{equation}\label{eq:multiperiod}
\begin{split}
\mu_{\widetilde Y} = \mu^{n}_{Y},\qquad \omega^2_{\widetilde Y} = \omega^{2n}_{Y},\qquad 
\HR^2_{\widetilde Y} = \left(\frac{\mu_Y^2}{\omega_Y^2}\right)^{2n},\\
1-\HR^2_{\widetilde X} - \HR^2_{\widetilde Y} =  \left(1- \HR^2_X - \HR^2_{Y}\right)\sum_{t=0}^{n-1}\HR^{2t}_Y.
\end{split}
\end{equation}
Hence, the knowledge of $\mu_Y$, $\omega_Y^2$, and $\HR_X^2$ determines the values of $\mu_{\widetilde Y}$, $\omega_{\widetilde Y}^2$, and $\HR_{\widetilde X}^2$ in a multiperiod model with IID returns.

By Subsection~\ref{SS:3}, the unconditional efficient frontier in the $(\mu,\omega)$--space reads
$$\omega^2 = \omega_{\widetilde Y}^2 +\HR_{\widetilde X}^{-2} (\mu-\mu_{\widetilde Y})^2.$$
If desired, one can now calculate the values of $\mu_{\widetilde Z}$, $\sigma_{\widetilde Z}^2$ via \eqref{mu_Z}--\eqref{sig_Z}. On converting $\HR_{\widetilde X}^2$ to $\SR_{\widetilde X}^2$ by means of \eqref{eq:HRSR}, one obtains the unconditional efficient frontier in the $(\mu,\sigma)$--space,
$$\sigma^2 = \sigma_{\widetilde Z}^2 + \SR_{\widetilde X}^{-2}(\mu-\mu_{\widetilde Z})^2.$$
A fully worked numerical example is presented in Appendix~\ref{S:A}. 

\section{Concluding remarks}\label{S:6}
The Hansen ratio arises naturally in the description of the mean--variance efficient frontier through its link to expected utility maximization. The latter takes on special significance in the dynamic setting. We have seen that the efficient frontier is fully described by any two of the three special portfolios $\widetilde X$, $\widetilde Y$, and $\widetilde Z$. Yet only the computation of $\widetilde X$ and $\widetilde Y$ is time-consistent in a dynamic setting. This points to a dichotomy in an effective evaluation of the efficient frontier.
When a risk-free asset is assumed, $\widetilde Z$ is known a-priori and only $\widetilde Y$ needs to be calculated. This is the situation in \citet{cerny.kallsen.07}. The much more difficult case when $\widetilde Z$ is to be calculated in a dynamic setting has received very limited attention in the literature; see \citet{li.ng.00} and \citet{lim.04,lim.05}. The approach proposed here, namely computing $\widetilde X$ and $\widetilde Y$ (or instead of $\widetilde X$ a suitable mix, such as $\widetilde X+\frac{\mu_{\widetilde Y}}{\omega^2_{\widetilde Y}}\widetilde Y$), leads to substantial simplification in such circumstances as illustrated in Section~\ref{S:5}.

In this paper, the Hansen ratio emerges as a long lost twin of the Sharpe ratio. That it has remained hidden for so long is likely due to an early disconnection between the mean--variance analysis on the one hand and the expected utility maximization on the other. The missing link was finally uncovered in \citet{filipovic.kupper.07} who characterize the mean--variance utility as the cash-invariant hull of the expected quadratic utility. This is what ties the Hansen ratio and the Sharpe ratio inextricably together and explains why the Hansen ratio is to orthogonal investments what the Sharpe ratio is to uncorrelated investments. Both quantities are clearly fundamental. It may have been hidden for a long time but the Hansen ratio is here to stay.

\appendix
\section{Efficient frontier in a multiperiod model}\label{S:A}
The numerical values in this example are based on \citet[Example~1]{li.ng.00}.
There are 3 risky assets with IID one-period total returns whose mean and variance, respectively, are given by 
\[
\mu_R =\left( 
\begin{array}{c}
1.162 \\ 
1.246 \\ 
1.228%
\end{array}%
\right) ,\qquad
\Sigma_R =\left( 
\begin{array}{ccc}
0.0146 & 0.0187 & 0.0145 \\ 
0.0187 & 0.0854 & 0.0104 \\ 
0.0145 & 0.0104 & 0.0289%
\end{array}%
\right).
\]%
This yields the second (co)moment matrix 
\[
\Omega_R = \Sigma_R +\mu_R \mu_R ^{\top
}=\left( 
\begin{array}{ccc}
                  1\,364\,844 &           1\,466\,552  &           1\,441\,436\\
                  1\,466\,552 &           1\,637\,916  &           1\,540\,488\\
                  1\,441\,436 &           1\,540\,488  &           1\,536\,884
\end{array}%
\right)\times 10^{-6}.
\]%

We have 
$$\HR_X^2+\HR_Y^2 = \mu_R^\top \Omega_R^{-1} \mu_R = \frac{28\,147\,713\,781}{28\,448\,540\,506}\approx 0.98943.$$
Furthermore,
\begin{align*}
\frac{\mu_Y}{\omega_Y^2} &{}= \mathbf{1}^\top\Omega_R^{-1}\mu_R = \frac{12\,123\,548\,000}{14\,224\,270\,253}\approx 0.85231;\\
\omega_Y^2 &{}= \frac{1}{\mathbf{1}^\top\Omega_R^{-1}\mathbf{1}} =\frac{14\,224\,270\,253}{16\,329\,740\,000}\approx 0.87107,
\end{align*}
yielding 
$$\HR_{Y}^{2} =\frac{\mu _{Y}^{2}}{\omega _{Y}^{2}}=\frac{7\,349\,020\,805\,415\,200}{11\,613\,931\,746\,061\,211}\approx 0.63278, $$
together with
\begin{align*}
\HR_{X}^{2}&{}=\mu_R^\top \Omega_R^{-1} \mu_R - \frac{(\mathbf{1}^\top\Omega_R^{-1}\mu_R)^2}{\mathbf{1}^\top\Omega_R^{-1}\mathbf{1}}
=\frac{582\,399}{1\,632\,974}\approx 0.35665,\\
\mu _{Y}&{}=\frac{\mathbf{1}^\top\Omega_R^{-1}\mu_R}{\mathbf{1}^\top\Omega_R^{-1}\mathbf{1}}=\frac{3\,030\,887}{4\,082\,435}
\approx 0.74242.
\end{align*}%

In a dynamic setting with $n=4$ periods one obtains from \eqref{eq:multiperiod}
\begin{align*}
{\HR}_{\widetilde X}^{2} &{}=\frac{1-\HR_{Y}^{2n}}{1-\HR_{Y}^{2}}\HR_{X}^{2}\approx 0.\,815\,50,\\
{\mu }_{\widetilde Y} &{}=\mu _{Y}^{n}=\left( \frac{3030\,887}{4082\,435}\right) ^{4}\approx 0.\,303\,81, \\
{\omega }_{\widetilde Y}^{2} &{}=\omega _{Y}^{2n}=\left( \frac{14\,224\,270\,253}{16\,329\,740\,000}\right)^{4}\approx 0.\,575\,71.
\end{align*}%
The efficient frontier in $(\mu,\omega)$--space then reads
\begin{align*}
\omega^{2} &{}={\omega }_{\widetilde Y}^{2}+{\HR}_{\widetilde X}^{-2}\left( \mu-{\mu }_{\widetilde Y}\right) ^{2} 
\approx 0.\,575\,71+1.\,226\,25(\widetilde{\mu }-0.\,303\,81)^{2}.
\end{align*}

One may alternatively choose to evaluate the parameters of the minimum variance portfolio as shown in Subsection~\ref{SS:Z}, 
\begin{align*}
{\mu }_{\widetilde Z} &{}=\frac{{\mu }_{\widetilde Y}}{1-{\HR}_{\widetilde X}^{2}}\approx 1.\,646\,63,\\
{\sigma }_{\widetilde Z}^{2} &{}=\frac{{\omega }_{\widetilde Y}^{2}}{1-{\HR}_{\widetilde X}^{2}}
\left( 1-{\HR}_{\widetilde Y}^{2}-{\HR}_{\widetilde X}^{2}\right) \approx 0.\,075\,446,\\
{\SR}_{\widetilde X}^{-2} &{}={\HR}_{\widetilde X}^{-2}-1\approx 0.\,226\,25,
\end{align*}
which yields the efficient frontier in the $(\mu,\sigma)$--space,
\begin{align*}
\widetilde{\sigma }^{2} &{}={\sigma }_{\widetilde Z}^{2}+\SR_{\widetilde X}^{-2}\left( \mu-{\mu }_{\widetilde Z}\right) ^{2} 
\approx 0.\,075\,446+0.\,226\,25(\widetilde{\mu }-1.\,646\,63)^{2}.
\end{align*}

All numerical values shown here are precise to the last digit, subject to rounding. Note, however, that the value od $\mu_{\widetilde Z}$ in \citet[p.~403]{li.ng.00} has a small rounding error.

\end{document}